\documentstyle[12pt]{amsart}
\title[\tiny From Agmon-Kannai expansion to $\hbox{\tiny KdV}$ hierarchy]
{From Agmon-Kannai expansion 
to \\ Korteweg-de Vries hierarchy}
\author[\tiny Iosif Polterovich]{Iosif Polterovich}
\address{Department of Theoretical Mathematics, The Weizmann Institute
of Science, Rehovot, Israel}
\email{iossif@@wisdom.weizmann.ac.il}
\def \phi{\varphi}
\def \epsilon{\varepsilon}
\numberwithin{equation}{subsection}
\theoremstyle{definition}

\theoremstyle{plain}

\newtheorem{theorem}[equation]{Theorem}

\begin{document}
\maketitle
\begin{abstract}
We present a new method for computation of the Korteweg--de Vries hierarchy 
via heat invariants of the $1$-dimensional Schr\"odinger operator.
As a result new explicit formulas for the KdV hierarchy are obtained. 
Our method is based on an asymptotic expansion of 
resolvent kernels of elliptic operators  due to S.~Agmon and Y.~Kannai.
\end{abstract}
\section{Introduction and main results}
\subsection{Heat asymptotics of Schr\"odinger operator and the KdV hierarchy}
Consider the $1$-dimensional Schr\"odinger (or Sturm-Liouville) operator
\begin{equation}
\label{schr}
L=\frac{\partial^2}{\partial x^2}+u(x).
\end{equation}
Its heat kernel $H(t,x,y)$ is  
a fundamental solution of the heat equation 
$$\left(\frac{\partial}{\partial t} - L\right)f=0.$$
The heat kernel has the following  asymptotics 
on the diagonal as $t\to~0+$ (see [3]): 
\begin{equation}
\label{ae}
H(t,x,x)\sim\frac{1}{\sqrt{4\pi t}}\sum_{n=0}^{\infty}h_n[u]t^n,
\end{equation}
where $h_n[u]$ are some polynomials in $u(x)$ and 
its derivatives. The coefficients $h_n[u]$ are called {\it heat invariants}
of the $1$-dimensional Schr\"odinger operator. 

Computation of heat invariants of self-adjoint elliptic operators
is a well-known problem in spectral theory which has many applications,
in particular to geometry and theoretical physics (see [3, 4, 6]).
Heat asymptotics of the $1$-dimensional Schr\"odinger operator
are of particular interest due to their relation to the
{\it Korteweg-de Vries (KdV) hierarchy} 
which is one of the basic objects in the theory of integrable systems 
(see [9, 10]). Namely, the KdV hierarchy is defined by ([2]):
\begin{equation}
\label{kdvh}
\frac{\partial u}{\partial t}=\frac{\partial}{\partial x}G_n[u],
\end{equation}
where
\begin{equation}
\label{q}
G_n[u]=
\frac{(2n)!}{2\cdot n!}h_n[u], 
\quad n\in {\Bbb N}.
\end{equation}
Set $u_0=u$, $u_n=\partial^n u/\partial x^n$, $n\in {\Bbb N}$,
where $u_n$, $n\ge 0$ are formal variables. 
The sequence of polynomials $G_n[u]=G_n[u_0,u_1,u_2,\dots]$ 
starts with~([2]): 
\begin{equation}
\label{ex}
G_1[u]=u_0, \,\, G_2[u]=u_2+3u_0^2, \,\, 
G_3[u]=u_4+10u_0u_2+5u_1^2+10u_0^3, \, \dots
\end{equation}
In particular, substituting $G_2[u]$ into (\ref{kdvh}) 
we obtain the familiar Korteweg-de Vries  equation ([9, 10, 5, 2]):
\begin{equation*}
\frac{\partial u}{\partial t}
=\frac{\partial^3 u}{\partial x^3}
+6u\frac{\partial u}{\partial x}.
\end{equation*}
\subsection{Explicit formulas for the KdV hierarchy and the heat invariants}
There are many ways to compute the polynomials $G_n[u]$ and $h_n[u]$. 
Originally it was done through a recursive 
relation between the heat invariants (see [5, 8]). 
However, in order to understand the structure of heat invariants 
and the KdV hierarchy closed expressions are desirable. 
This might also make actual computations easier.
Such expressions were first obtained by 
R.~Schimming and I.~Avramidi ([13, 2]), all of them having a very 
complicated combinatorial structure. 

In this paper we present a method for computation 
of heat invariants of the $1$-dimensional Schr\"odinger operator 
(see [11] for a similar method for the Laplacian 
on a $2$-dimensional Riemannian manifold).
As a result we obtain simpler explicit formulas for $h_n[u]$ and hence for 
the KdV hierarchy. 
Our  method is based on an asymptotic expansion of resolvent kernels of 
elliptic operators due to S.~Agmon and Y.~Kannai ([1, 11]).

We introduce the following notation.
Consider $L^k \lbrack x^{2k-2n} \rbrack|_{x=0}$, $k\ge n\ge 1.$
This is a polynomial in $u(0), u'(0),\dots,u^{(2k-2)}(0)$.
Let us make a formal change of variables: 
$u^{(i)}(0)\to u_i$, $i=0,\dots,2k-2$,
and denote the obtained polynomial by $P_{kn}[u]$.  

\begin{theorem}
\label{KdV}
The polynomials $G_n[u]$, $n \in {\Bbb N}$,
are given by:
$$
G_n[u]=
\sum_{m=n}^{2n}\sum_{k=n}^m
\frac{(-1)^{n+k}\binom{m}{k}\binom{m}{n}\binom{2m}{m}}
{2^{2m-2n+1}\binom{2m}{2n} (2k-2n)!} 
P_{kn}[u],
\eqno (\ref{KdV})
$$
\end{theorem}

\medskip

Expanding the expression for $L^k$ (see [12]) we  get:
\begin{theorem}
\label{KdV1}
The  polynomials $G_n[u]$, $n\in {\Bbb N}$ are equal to:
$$
G_n[u]=\sum_{m=n}^{2n}\sum_{k=n}^m
\frac{(-1)^{n+k}\binom{m}{k}\binom{m}{n}\binom{2m}{m}}
{\binom{2m}{2n}2^{2m-2n+1}}
\sum_{p=1}^k\mskip-2\thinmuskip
\sum_{j_1,\dots,j_p\atop j_1+\cdots +j_p=2(n-p)}\mskip-13\thinmuskip
C_{j_1,\dots,j_p}u_{j_1}\cdots u_{j_p},
\eqno(\ref{KdV1})
$$
where 
\begin{equation*}
C_{j_1,\dots,j_p}=\sum\begin{Sb}
0\le\l_0\le l_1\le \cdots\le l_{p-1}=k-p\\ 
2l_i\ge j_1+\cdots+j_{i+1},\,\,i=0,\dots,p-1.\end{Sb}
\binom{2l_0}{j_1}\binom{2l_1-j_1}{j_2}\cdots
\binom{2l_{p-1}-j_1-\cdots-j_{p-1}}{j_p}.
\end{equation*}
\end{theorem}      

Expressions for the heat invariants $h_n[u]$ can be easily deduced
from (\ref{KdV}) and (\ref{KdV1}) using (\ref{q}).

Theorems \ref{KdV} and \ref{KdV1} are proved in sections 3.1 and 3.2.

\section{Computation of heat invariants}
\subsection{A modification of Agmon-Kannai expansion}
The original Agmon--Kan\-nai theorem ([1]) deals with asymptotic behaviour
of resolvent kernels of elliptic operators. In [11] we have obtained
a concise reformulation of this theorem suitable
for computation of heat invariants. Before presenting it we introduce
some notations.

Let $H$ be a a self--adjoint elliptic differential 
operator of order $s$ on a Riemannian manifold $M$ of dimension $d<s$ 
and let $H_0$ be the operator obtained by
freezing the coefficients of the principal part $H'$ of the operator
$H$ at some point $x\in M$: $H_0=H'(x)$.  
Denote by $R_\lambda(x,y)$ the kernel of the resolvent
$R_\lambda=(H-\lambda)^{-1}$, and by $F_{\lambda}(x,y)$ --- the 
kernel of $F_{\lambda}=(H_0-\lambda)^{-1}$.
\begin{theorem}$\operatorname([11]).$
\label{ak}
The resolvent kernel $R_{\lambda}(x,y)$ 
has the following asymptotic representation on the diagonal
as $\lambda \to \infty$:
\begin{equation}
\label{s}
R_\lambda(x,x) \sim \rho(x)^{-1} \sum_{m=0}^\infty X_m F_\lambda^{m+1}(x,x),
\end{equation}
where $\rho(x)dx$ is the volume form on the manifold $M$ and 
the operators $X_m$ are defined by:
\begin{equation}
\label{y}
X_m=\sum_{k=0}^m (-1)^k \binom{m}{k} H^k H_0^{m-k}, \,\, m\ge 0.
\end{equation}
\end{theorem}
\subsection{From the resolvent to the heat kernel} 
Let us return to our particular case of $1$-dimensional Schr\"odinger operator.
We have 
\begin{equation*}
M={\Bbb R}^1, \quad
d=1,  \quad
s=2, \quad
H \equiv -L, \quad H_0 \equiv -\frac{d^2}{dx^2}, \quad
\rho(x)\equiv 1.
\end{equation*}
Note that (\ref{s}) is an expansion in powers of $-\lambda$ due to the 
following formula ([1]):
\begin{equation}
\label{int}
\frac{d^{\alpha}}{dx^{\alpha}}
F_\lambda^{m+1}(x,x)=
(-\lambda)^{\frac{\alpha-2m-1}{2}}\,\frac{(-1)^{\frac{\alpha}{2}}}{2\pi}
\int_{-\infty}^{\infty}
\frac{\xi^{\alpha}d\xi}
{(\xi^2+1)^{m+1}}.
\end{equation}
Collecting terms with the same powers of $-\lambda$ in (\ref{s}) 
one receives the standard asymptotic expansion of the kernel  of
$R_{\lambda}=(-L-\lambda)^{-1}$ on the diagonal as $\lambda \to \infty$ 
(see [1]):
\begin{equation}
\label{x}
R_{\lambda}(x,x) \sim (-\lambda)^{-1/2} \sum_{n=0}^{\infty}r_n[u]
(-\lambda)^{-n/2},
\end{equation}
where $r_n[u]$ are polynomials in $u(x)$ and its derivatives.

For each $n\in {\Bbb N}$ the polynomials $r_n[u]$ and the 
heat invariants $h_n[u]$ differ by a constant multiplier (cf. [11]):
\begin{equation}
\label{bn}
r_n[u]=\frac{(2n)!}{2^{2n+1}n!}h_n[u],
\end{equation}
\section{Proofs of main theorems}

\subsection{Proof of Theorem \ref{KdV}.}
Due to (\ref{q}) and (\ref{bn}), 
\begin{equation}
\label{rn}
G_n[u]=2^{2n}r_n[u].
\end{equation}
Therefore in order to compute $G_n[u]$ we need to find 
the coefficient $r_n[u]$ in (\ref{x}).

Without loss of generality set $x=0$. We need  to collect all
terms in the sum (\ref{s})) containing $(-\lambda)^{-n/2}$. 
Let us apply formula (\ref{int}). 
Note that if $\alpha$ is odd then (\ref{int}) is zero identically.
For the terms we are interested in we have
$\alpha=2m-2n$ and hence $m\ge n$. On the other hand
the order of the differential operator $X_m$ defined by (\ref{y}) 
is not greater than $m$ (see Lemma 5.1, [1]).
Hence $\alpha=2m-2n\le m$ which implies $m\le 2n$.

Denote by $r'_n$ a polynomial in $u(0), u'(0), u''(0),...$ which
is obtained from the polynomial $r_n[u]$ by a formal change of variables 
$u_i\to u^{(i)}(0)$, $i\ge~0$.

Then due to (\ref{y}) we have:
\begin{equation*}
r'_n=\sum_{m=n}^{2n}\sum_{k=n}^m
(-1)^{m+k}\binom{m}{k}\Phi_{mn}L^k\frac{d^{2m-2k}}{dx^{2m-2k}}
\left.\left[\frac{x^{2m-2n}}{(2m-2n)!}\right]\right|_{x=0},
\end{equation*}
where
\begin{equation*}
\Phi_{mn}= \frac{(-1)^{m-n}}{2\pi}
\int_{-\infty}^{\infty}
\frac{\xi^{2m-2n}d\xi}
{(\xi^2+1)^{m+1}}=\frac{(-1)^{(m-n)}}{2\pi}B\left(m-n+\frac{1}{2};
n+\frac{1}{2}\right),
\end{equation*}
and the integral above is computed using the tables ([7]).

\smallskip

Let us rewrite the $B$-function in terms of factorials:
\begin{equation*}
\label{beta}
B\left(m-n+\frac{1}{2};
n+\frac{1}{2}\right)=\frac{\Gamma\left(m-n+\frac{1}{2}\right)
\Gamma\left(n+\frac{1}{2}\right)}{\Gamma(m+1)}=
\frac{(2n)!\cdot (2m-2n)!\cdot\pi}{2^{2m} \cdot m! \cdot n! \cdot (m-n)!}.
\end{equation*}

Substituting this into the expression for $\Phi_{mn}$ and simplifying $r_n'$ 
we finally get:
\begin{equation}
\label{xx}
r_n[u]=\sum_{m=n}^{2n}\sum_{k=n}^m
\frac{(-1)^{n+k}\binom{m}{k}\binom{m}{n}\binom{2m}{m}}
{2^{2m+1}\binom{2m}{2n} (2k-2n)!}
P_{kn}[u],
\end{equation}
where $P_{kn}[u]=P_{kn}[u_0, u_1,..,u_{2k-2}]$ are polynomials 
obtained from $L^k [x^{2k-2n}]|_{x=0}$  by a formal change 
of variables $$u(0)\to u_0, u'(0)\to u_1,\dots,u^{(2k-2)}(0)\to u_{2k-2}.$$
By (\ref{rn}), formula (\ref{xx}) implies (\ref{KdV}) and this
completes the proof of the theorem.
\qed

\subsection*{Remarks}
The proof of Theorem \ref{KdV} is similar to the
proof of Theorem 1.4 in [11]. Formula (\ref{KdV}) was programmed using 
Mathematica  ([14]) and for $1\le n\le 5$  the results
agreed with the already known ones (cf. (\ref{ex}),  [5]).

\subsection{\bf Proof of Theorem \ref{KdV1}.}
In [12] an explicit formula for the powers of the $1$-dimensional
Schr\"odinger operator was obtained. We may rewrite it in the form:
\begin{equation*}
L^k=\frac{d^{2k}}{dx^{2k}}+\sum_{p=1}^k\mskip-2\thinmuskip
\sum_{j_0,j_1,\dots,j_p\atop j_0+j_1+\cdots +j_p=2(k-p)}\mskip-13\thinmuskip
C_{j_1,\dots,j_p}u_{j_1}(x)..u_{j_p}(x)\frac{d^{j_0}}{dx^{j_0}},
\end{equation*}
where 
\begin{equation*}
C_{j_1,\dots,j_p}=\sum\begin{Sb}
0\le\l_0\le l_1\le \cdots\le l_{p-1}=k-p\\ 
2l_i\ge j_1+\cdots+j_{i+1},\,\,i=0,\dots,p-1.\end{Sb}
\binom{2l_0}{j_1}\binom{2l_1-j_1}{j_2}\cdots
\binom{2l_{p-1}-j_1-\cdots-j_{p-1}}{j_p}.
\end{equation*}

Setting $j_0=2k-2n$ this allows to expand the expression 
for $L^k [x^{2k-2n}]|_{x=0}$ and hence for the polynomials $P_{kn}[u]$.
Substituting it into (\ref{xx}) we immediately 
receive (\ref{KdV1}) which completes the proof of the theorem. 
\qed

\smallskip

\noindent {\bf Acknowledgments}. This paper is a part of my Ph.D. research
at the Department of Mathematics of the Weizmann Institute of Science. 
I am very grateful to my Ph.D. advisor Professor Yakar Kannai for constant  
support and valuable remarks.
\smallskip

\section*{References}

\bigskip

\noindent 1. S.Agmon and Y.Kannai, {\it On the asymptotic behavior of spectral 
functions and resolvent kernels of elliptic operators}, Israel J. Math. {\bf 5}
(1967), 1-30.

\smallskip

\noindent 2. I.V. Avramidi and R. Schimming, {\it A new explicit expression
for the Korteweg - de Vries hierarchy}, solv-int/9710009,  (1997), 1-17.

\smallskip

\noindent 3. M.~Berger, {\it Geometry of the spectrum}, Proc. Symp. Pure Math. 
{\bf 27} (1975), 129-152.

\smallskip

\noindent 4. S.A. Fulling ed., {\it Heat Kernel Techniques and Quantum Gravity},
Discourses in Math. and its Appl., No. 4, Texas A\&M Univ., 1995.

\smallskip

\noindent 5.  I.M. Gelfand, L.A.Dikii, {\it Asymptotic behaviour of the
resolvent of Sturm-Liouville equation and the algebra of the 
Korteweg-de Vries equations}, Russian Math. Surveys, {\bf 30:5} (1975),
77-113. 

\smallskip

\noindent 6. P.~Gilkey, {\it Heat equation asymptotics}, Proc. Symp. 
Pure Math. {\bf 54} (1993), 317-326. 

\smallskip

\noindent 7. I.S. Gradstein, I.M. Ryzhik, {\it Table of integrals, series and
products}, Academic Press, 1980.

\smallskip

\noindent 8. H.P. McKean, P. van Moerbeke, {\it The spectrum of Hill's 
equation}, Invent. Math.,  {\bf 30} (1975), 217-274.

\smallskip

\noindent 9.  S.P. Novikov , {\it Role of integrable models in the development
of mathematics}, in: Lectures of Fields medalists, M.~Atiyah and
D.~Iagolnitzer eds., World Scientific, Singapure University Press (1997),
202-218.

\smallskip

\noindent 10. S.P. Novikov, S.V. Manakov, L.P. Pitaevskii, V.E. Zakharov,
{\it Theory of solitons: the inverse scattering method}, 
Consultants Bureau, 1984.

\smallskip

\noindent 11. I.~Polterovich, {\it Commutator method for computation of heat 
invariants}, math.dg/ 9805047, (1998), 1-11.

\smallskip

\noindent 12.  S.Z. Rida, {\it Explicit formulae for the powers of a 
Schr\"odinger-like
ordinary differential operator}, J. Phys. A: Math. Gen., {\bf 31} (1998), 5577-5583.

\smallskip

\noindent 13. R. Schimming, {\it An explicit expression for the 
Korteweg-de Vries hierarchy}, Acta Appl. Math., {\bf 39} (1995), 489-505.

\smallskip

\noindent 14.  S.Wolfram, {\it Mathematica: a system for doing mathematics 
by computer}, Addison-Wesley, 1991.

\end{document}